\documentclass[prb,twocolumn]{revtex4}
\usepackage{epsf}

\begin{document}

\title{Shot noise in diffusive ferromagnetic metals}
\author{M. Hatami and M. Zareyan}

\affiliation{Institute for Advanced Studies in Basic Sciences,
45195-1159, Zanjan, Iran}

\begin{abstract}
We show that shot noise in a diffusive ferromagnetic wire
connected by tunnel contacts to two ferromagnetic electrodes can
probe the intrinsic density of states and the extrinsic impurity
scattering spin-polarization contributions in the polarization of
the wire conductivity.  The effect is more pronounced when the
electrodes are perfectly polarized in opposite directions. While
in this case the shot noise has a weak dependence on the impurity
scattering polarization, it is strongly affected by the
polarization of the density of states. For a finite spin-flip
scattering rate the shot noise increases well above the normal
state value and can reach the full Poissonian value when the
density of states tends to be perfectly polarized. For the
parallel configuration we find that the shot noise depends on the
relative sign of the intrinsic and the extrinsic polarizations.
\end{abstract}

\pacs{ 74.40.+k, 75.47.-m, 72.25.Rb, 73.23.-b} \maketitle

Shot noise is the low temperature temporal fluctuations of the
electrical current through a conducting structure caused by the
randomness of the electron scattering and the Fermi statistics. In
the past years the shot noise has been extensively studied in
different types of mesoscopic structures \cite{Blanter00}. It has
been revealed that the shot noise measurements provide valuable
information about the charge transport process which are not
extractable from the mean conductance. For a fully random
transmission of the electrons ( e. g., through a tunnel contact)
the current fluctuations $\Delta I(t)$ around the mean current
$\bar I$ is described by a Poissonian noise power $S=2e \bar I$
\cite{Schottky}.
\par
Correlations can reduce the shot noise below the Poissonian value.
In a diffusive metal the restriction imposed by the Pauli
exclusion principle on the random scattering of electrons from the
impurities reduces the noise power by a universal factor of
one-third \cite{Beenakker92,Steinbach96}. Coulomb repulsion
introduces another source of the correlations which changes the
shot noise in the diffusive metals \cite{Steinbach96,Nagaev95} as
well as a nondegenerate electron gas \cite{Gonzalez98}. Probing
correlations and interactions by the shot noise measurement has
been the subject of many investigations in the recent years
\cite{Blanter00,Sauret04}. In contrary effects of spin-dependent
correlations on the current fluctuations through a diffusive
conductor has received little attention. In particular a natural
question is that what happens to the one-third shot noise in a
diffusive {\it ferromagnetic} metal in which additional
correlations can be introduced between carriers of opposite spins
by the interplay between the spin polarization of the density of
states (DOS) of the Fermi level and the spin-dependent
scatterings.
\par
Here we propose that the shot noise in ferromagnetic metals can
probe such correlations between spin up and down electrons. The
basic idea is that the spin-dependent scattering including the
spin-orbit coupling and the normal and the magnetic impurity
scatterings are correlated with the imbalance in the Fermi level
DOS of opposite spins. We show that such correlations change the
current fluctuations in a way that shot noise acquires different
dependence and sensitivity on the {\it intrinsic} spin-dependence
due to the polarization of the conduction band, and the {\it
extrinsic} spin polarization induced by the impurity scattering.
Our finding is important in the context of the anomalous Hall
effect \cite{Hall880}. In ferromagnets the Hall resistance
contains an anomalous term that is proportional not to the
external field but to the magnetization of the ferromagnet. The
question whether the anomalous Hall effect originates from the
intrinsic \cite{Karplus54} spin-polarization of the band structure
or has an extrinsic \cite{Smit55} origin, such as scattering by
normal or magnetic disorder, has been a subject of controversy
\cite{Kotzler05}. Thus the shot noise measurement in ferromagnets
can be proposed to probe the intrinsic and the extrinsic anomalous
Hall resistances.
\par
Our motivation also comes from the importance of the noise in
spintronics \cite{Zutic04} devices in view of applications.
Recently spin-polarized shot noise has been studied theoretically
in normal metals connected by the ferromagnetic terminals with
collinear \cite{Mishchenko03,Zareyan051} and noncollinear
\cite{Tserkovnyak01} magnetization directions. In Ref.
\cite{Zareyan051} a semiclassical Boltzmann-Langevin theory of the
spin-polarized current fluctuations in diffusive normal metals was
developed. It was found that in a multi-terminal spin-valve
structure the shot noise and the cross correlations measured
between currents of two different ferromagnetic terminals can
deviate substantially from the unpolarized values, depending on
the relative orientation of the magnetizations, the degree of
spin-polarization of the terminals and the strength of the
spin-flip scattering in the normal conductor.
\par
\begin{figure}
\vspace{0.2in} \centerline{\hbox{\epsfxsize=2.7in
\epsffile{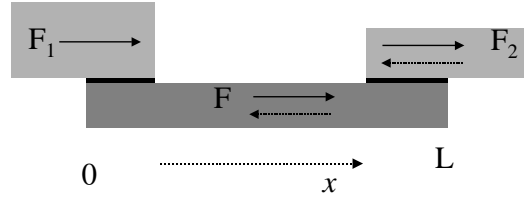}}} \caption{Schematic of the studied
spin-valve structure.} \label{hzfig1}
\end{figure}
In this paper we study fluctuations of the electrical current
through a diffusive ferromagnetic wire.  We consider a full
ferromagnetic spin-valve system as shown in Fig. \ref{hzfig1}. A
diffusive ferromagnetic wire is connected through the tunnel
contacts to two ferromagnetic reservoirs. The reservoirs F$_{1}$
and F$_{2}$ are held at equilibrium at the voltages $0$ and $V$
respectively. Each of the F-terminals and the connecting wire has
a magnetization which can be pointed in two different collinear
directions. The polarization of the electronic DOS in the
connected F metal causes a spin-polarized tunneling through the
tunnel contact which is characterized by a spin-dependent
conductance. For spin conserving tunneling the conductance of the
$i$th ($i=1,2$) contact  has the form $g_{i\alpha}=g_{Ti}(1+\alpha
p_{N})(1+\alpha p_{i})/2$. Here $g_{Ti}$ is the tunnel conductance
in a fully normal system, $p_i$ the polarization of the $i$th
terminal, $p_{N} =\sum_{\alpha}\alpha
N_{\alpha}/\sum_{\alpha}N_{\alpha}$ the DOS polarization with
$N_{\alpha}$ being the spin $\alpha$ DOS in the F-wire
($\alpha=\pm 1$ denote the spin indices).
\par
In the semiclassical regime the electronic transport across the
F-wire can be explained within an extension of the
Boltzmann-Langevin theory, which is developed in Refs.
\cite{Mishchenko03,Zareyan051} to study the spin-polarized shot
noise in normal metals. We generalize this theory to a
ferromagnetic metal by including the spin-dependence of the
conductivity and the spin-flip diffusion length. Here we proceed
with the corresponding diffusion equations for the fluctuating
current density $
j_{\alpha}(x,t,\varepsilon)=\overline{j}_{\alpha}(x,\varepsilon)+
\delta j_{\alpha}(x,t,\varepsilon)$ and the distribution function
$f_{\alpha}(x,t,\varepsilon)=\overline{f}_{\alpha}(x,\varepsilon)+
\delta f_{\alpha}(x,t,\varepsilon)$ of spin $\alpha$ electrons at
the energy $\varepsilon$, which have the form
\begin{eqnarray}
\label{jalpha} &&j_{\alpha}= -\sigma_{\alpha}
\frac{\partial}{\partial x}f_{\alpha} + j^{{\text c}}_{\alpha}, \\
 \label{divjalpha}
&&\frac{\partial}{\partial x}
j_{\alpha}=-\frac{\sigma_{\alpha}}{2\ell_{{\text
{sf}}\alpha}^{2}}( f_{\alpha}-f_{-\alpha})+ i^{{\text
{sf}}}_{\alpha},
\end{eqnarray}
where the spin-dependent conductivity in the F-wire
$\sigma_{\alpha}=e^{2} N_{\alpha}D_{\alpha}$, in which
$D_{\alpha}=v^2_{{\text F}\alpha}\tau_{\alpha}/3$ is the spin
$\alpha$ diffusion constant . The spin-dependent relaxation time
$\tau_{\alpha}$ is expressed in terms of the normal impurity
$\tau_{\alpha}^{{\text imp}}$ and the spin-flip scattering
$\tau_{\alpha}^{{\text {sf}}}$ relaxation times as
$1/\tau_{\alpha}=1/\tau_{\alpha}^{{\text
imp}}+1/2\tau_{\alpha}^{{\text {sf}}}$. The spin-flip diffusion
length $\ell_{{\text {sf}}}$ is expressed as $1/\ell_{{\text
{sf}}}^{2}=1/2\ell^{2}_{{\text {sf}}+}+1/2\ell^{2}_{{\text
{sf}}-}$ with $\ell_{{\text {sf}}\alpha}=(D_{\alpha}\tau^{{\text
{sf}}}_{\alpha})^{1/2}$. In the diffusive limit of $\ell_{{\text
{imp}}}\ll L$ we will consider the more realistic case where
$\ell_{{\text {sf}}}$ is much larger than $\ell_{\text {imp}}$,
but arbitrary compared to $L$.
\par
In Eq. (\ref{jalpha})  $j^{\text{c}}_{\alpha}$  is the Langevin
source of the current density fluctuations caused by the
stochastic nature of the normal and the spin-flip scattering
events. The spin is not conserved by the spin-flip scattering
which leads to the appearance of an additional fluctuating
divergent term $i^{{\text {sf}}}_{\alpha}$ in Eq.
(\ref{divjalpha}). We obtain the following results for the
correlations of the fluctuating terms in Eqs. (\ref{jalpha}) and
(\ref{divjalpha}),
\begin{eqnarray}
\label{jcjc} &&
<j^{{\text c}}_{\alpha}(x,t,\varepsilon)j^{{\text c}}_{\alpha'}(x',t',\varepsilon')>=\nonumber\\
&&\delta_{\alpha\alpha'}\Delta
\sigma_{_{\alpha}}[\frac{2\tau_{\alpha}}{\tau^{{\text
imp}}_{\alpha}} \Pi_{\alpha\alpha}+
\frac{\tau_{\alpha}}{2\tau^{{\text {sf}}}_{\alpha}}(\Pi_{\alpha-\alpha}+\Pi_{-\alpha\alpha})],\\
\nonumber\\
\label{isfisf} &&
<i^{{\text {sf}}}_{\alpha}(x,t,\varepsilon)i^{{\text {sf}}}_{\alpha'}(x',t',\varepsilon')>=\nonumber\\
&&(\delta_{\alpha\alpha'}-\delta_{-\alpha\alpha'}) \Delta
\frac{\sigma_{\alpha}}{2\ell_{{\text
sf}\alpha}^{2}}[\Pi_{\alpha-\alpha}+\Pi_{-\alpha\alpha}],
\end{eqnarray}
in which
$\Delta=\delta(x-x')\delta(t-t')\delta(\varepsilon-\varepsilon')$
and
$\Pi_{\alpha\alpha'}=\bar{f}_{\alpha}(x,\varepsilon)(1-\bar{f}_{\alpha'}(x',\varepsilon'))$.
Eqs. (\ref{jcjc}-\ref{isfisf}) are the extension of the
corresponding relations obtained for the normal transport
\cite{Blanter00,Beenakker92}. Thus we find that the current
correlations are determined by the mean distribution of the spin
$\alpha$ electrons.
\par
Equations for the mean distribution functions are deduced by
combining Eqs. (\ref{jalpha}) and (\ref{divjalpha}) as
\begin{eqnarray}
&&\frac{\partial^{2}}{\partial
x^{2}}\bar{f}_{\alpha}=\frac{1}{2\ell_{{\text {sf}}\alpha}^{2}}
(\bar{f}_{\alpha}-\bar{f}_{-\alpha}),
\end{eqnarray}
which for the two terminal structure of Fig. (\ref{hzfig1}) has
solutions of the form,
\begin{eqnarray}
&& \bar{f}_{\alpha}=f_{1}+(f_{2}-f_{1})
[a+b \frac{x}{L}\nonumber\\
&& +(\alpha-p_{\sigma})(c~ \sinh{\frac{\lambda x}{L}}+d~ \cosh{
\frac{\lambda x}{L}})]. \label{meanf}
\end{eqnarray}
Here $\lambda=L/\ell_{{\text {sf}}}$ measures the strength of the
spin-flip scattering which can be expressed in terms of the normal
state spin-flip strength $\lambda_0$ and the DOS polarization $p_
N$ as $\lambda= \lambda_0 (1-p_{ N}^{2}+\gamma(1+
p_{N}^{2}))^{1/2}$; $f_{i}=f_{\text{FD}}(\varepsilon-eV_{i})$ is
the Fermi-Dirac distribution function in the electrodes held at
the voltages $V_{i}$ and $p_{\sigma}=\sum_{\alpha} \alpha
\sigma_{\alpha}/\sum_{\alpha} \sigma_{\alpha}$ stands for the
polarization of the conductivity. Both the polarization of DOS as
well as the spin dependent scattering rate contribute to
$p_{\sigma}$. Disregarding the Fermi velocity polarization it can
be written as
\begin{eqnarray}
\label{psigma} &&p_{\sigma}=\frac{(1-p_{N}^{2})~p_{W}+2\gamma
~p_{N}}{1-p_{N}^{2}+ \gamma(1+p_{N}^{2})},
\end{eqnarray}
Here $p_ W=\sum_{\alpha}\alpha W^{{\text
imp}}_{\alpha}/\sum_{\alpha} W^{{\text imp}}_{\alpha}$ is the
polarization of the impurity scattering rate and $\gamma=2
W^{{\text {sf}}}_{0}/\sum_{\alpha} W^{{\text imp}}_{\alpha}\ll 1$
is the ratio between the spin-flip and the normal impurity
scattering rates.
\par
The unknown coefficients $a, b,c, d$ in Eq. (\ref{meanf}) are
obtained by the boundary conditions. Assuming spin-conserving
tunneling the boundary conditions are expressed as the
conservation of the temporal spin $\alpha$ current through the
contacts. The fluctuating spin $\alpha$ current in the contact $i$
is written as
\begin{equation}
I_{i \alpha}(\varepsilon,t)=g_{i \alpha} [ f_{i}-
f_{\alpha}(x_{i},t)]+ \delta I_{i \alpha},
 \label{IinT}
\end{equation}
where $\delta I_{i \alpha}$ are the intrinsic fluctuations of the
current due to the scattering of the electrons by the tunnel
barriers. The intrinsic current fluctuations have the correlations
of a Poissonion process of the form $\langle\delta I_{i\alpha}
\delta I_{j \alpha'}\rangle= \delta_{\alpha\alpha'} \delta_{ij} e
\bar I_{\alpha}$.
\par
On the other hand we can calculate the spin $\alpha$ current at a
given point $x$ from the diffusion equations (\ref{jalpha})and
(\ref{divjalpha}). The mean current is obtained from the
distribution function, given by Eq. (\ref{meanf}), via the
relation
\begin{eqnarray}
&&\bar{I}_{\alpha}(x,\varepsilon)= -g_{\alpha}
\frac{\partial}{\partial x} \bar f_{\alpha}(x,\varepsilon).
\label{IinF}
\end{eqnarray}
in which $g_{\alpha}=\sigma_{\alpha}A/L$ are the spin dependent
conductances of the F wire. For the fluctuations of the currents at
the points $0,L$ we obtain the results
\begin{eqnarray}
&&\Delta I_{\alpha}(0,L)= g_{\alpha} \sum_{\alpha'}
q_{\sigma\alpha'}(\delta f_{\alpha'}(L,0)-\delta f_{\alpha'}(0,L)) \nonumber \\
&& +\alpha g_{-\alpha} s(\lambda)q_{\sigma \alpha}
\sum_{\alpha'}\alpha'(\delta f_{\alpha'}(L,0)- \cosh{\lambda}~\delta
f_{\alpha'}(0,L)) \nonumber \\
&&+q_{\sigma\alpha} \delta {\cal  I}^{{\text c}}_{c}(0,L)
+\frac{\alpha}{2} \delta{\cal  I}^{{\text
c}}_{s}(0,L)+(\frac{1-2q_{\sigma\alpha}}{2})\delta{\cal
I}^{\text{c}}_{cs}(0,L), \nonumber \\
\label{deltai}
\end{eqnarray}
Here $q_{\sigma \alpha}=(1+\alpha p_{\sigma})/2$,
$s(\lambda)=\lambda/ \sinh {\lambda}$ and the Langevin fluctuating
currents $\delta {\cal I}^{{\text c}}_{c}$,~$\delta {\cal I}^{{\text
c}}_{s}$ and $ \delta {\cal I}^{\text{c}}_{cs}$ are given by
\begin{eqnarray}
\label{delicc} \delta {\cal I}^{{\text c}}_{c}(0,L)= A
\sum_{\alpha}\int dx
(i^{{\text {sf}}}_{\alpha}+ j^{{\text c}}_{\alpha}\frac{\partial}{\partial x})\phi_{c0,L},\\
\label{delics} \delta {\cal I}^{{\text c}}_{s}(0, L)=A\sum_{\alpha}
\alpha \int dx
(i^{{\text {sf}}}_{\alpha}+j^{{\text c}}_{\alpha}\frac{\partial}{\partial x})\phi_{s0, L},\\
\label{delic} \delta {\cal I}^{{\text c}}_{cs}(0,L)=A
\sum_{\alpha} \int dx (i^{{\text {sf}}}_{\alpha}+j^{{\text
c}}_{\alpha}\frac{\partial}{\partial x})\phi_{s0,L},
\end{eqnarray}
with $\phi_{c0}(x)=1-x/L,
\phi_{s0}(x)=\sinh{[\lambda(1-x/L)]}/\sinh{ \lambda},
 \phi_{cL}(x)=x/L,$ and $\phi_{sL}(x)=\sinh{[\lambda
x/L]}/\sinh{\lambda}$.
\par
Now we impose the current conservation rule at the contacts using
the expressions for the spin $\alpha$ currents given by Eqs.
(\ref{IinT}-\ref{deltai}), from which we obtain the coefficients
$a,b,c,d$ and the fluctuations of the spin $\alpha$ distributions
$\delta f_{\alpha}(0,L)$ at the contact points $0,L$. The
distribution fluctuations are expressed in terms of $\delta
I_{i\alpha}$, and the Langevin currents $\delta {\cal I}^{\text
c}_{c(s)}(0,L)$, and $\delta {\cal I}^{\text c}_{cs}(0,L)$. These
results together with the relations for the correlations of the
Langevin currents which can be calculated using  Eqs.
(\ref{delicc}-\ref{delic}) and (\ref{jcjc}-\ref{isfisf}), allows
us to calculate the correlation of the spin $\alpha$ electrons in
each of the contacts. Thus the correlations of the charge current
$S=\langle\Delta I \Delta I\rangle$ and the Fano factor
$F=S/2e{\bar I}$ are obtained. The expression for the Fano factor
is too lengthy to be written down here. For simplicity we consider
a symmetric double barrier structure for that
$g_{T1}=g_{T2}=g_{T}$. In the absence of any polarization in the
system our result reduces to Fano factor for a full normal metal
double barrier structure \cite{Blanter00}
\begin{equation}
F_ N=\frac{1}{3}~\frac{12+12~g+6~g^{2}+g^{3}}{(2+
g)^{3}},\label{ffn}
\end{equation}
where $g=g_{T}/g_{N}$ with $g_{N}$ being the normal conductance of
the wire. In the following we discuss the variation of the Fano
factor with respect to $F_N$ taking $g=1$ and for $\gamma\ll 1$.
\par We consider perfectly polarized F-terminals, when
the effect of spin-polarization induced by the terminals is most
pronounced. The magnetization vectors in the terminals can have
parallel and anti-parallel orientations. The anti-parallel (A)
case presents a single configuration. For the parallel (P) case
there are two configurations corresponding to the parallel (PP)
and antiparallel (PA) alignment of the polarization of the wire
with respect to the magnetizations in the terminals.
\begin{figure}
\centerline{\hbox{\epsfxsize=3.in \epsffile{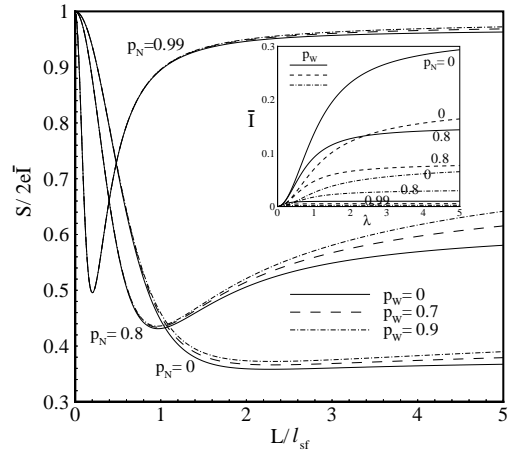}}}
\caption{Fano factor $F$ as a function of the spin-flip
  scattering strength $\lambda=L/\ell_{\text {sf}}$ for the antiparallel
  magnetization alignment of
  the electrodes. We parameterize the magnetic properties of the wire
  with the DOS polarization $p_N$ and the impurity scattering rate
  polarization $p_W$. Inset shows the corresponding mean current
  versus $\lambda$.} \label{hzfig2}
\end{figure}
\par
The most striking result is obtained for the A configuration. In
this case the perfectly antiparallel polarization of the terminals
prevent charge transport through the wire in the absence of the
spin-flip scattering. Thus one would expect more sensitive
dependence of the fluctuating current on the strength of the
spin-flip scattering which itself depends on the polarization of
the wire. Fig. \ref{hzfig2} shows the Fano factor $F$ and the mean
current ${\bar I}$ (inset) dependence on $\lambda$ for different
$p_N$ and $p_W$. The Fano factor is strongly modified with respect
to the normal value given in Eq. (\ref{ffn}). In the limit
$\lambda \rightarrow 0$ the mean current $I$ is vanishingly small
and the Fano factor takes its full Poissonion value, $1$. This
does not depend on the values of $p_N$ and $p_W$. The similar
effect was found before for the normal spin-valve system
\cite{Mishchenko03,Zareyan051}.
\begin{figure}
\centerline{\hbox{\epsfxsize=3.in \epsffile{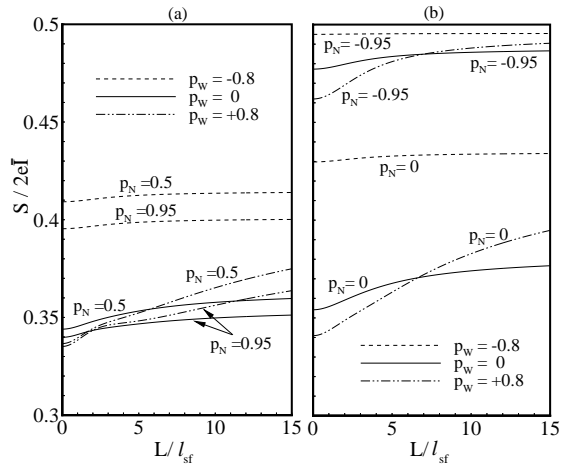}}}
\caption{Fano factor for parallel configuration of the electrodes
magnetizations when the magnetization of the wire is parallel (a)
and  antiparallel (b) to the electrodes magnetization.}
\label{hzfig3}
\end{figure}
We can understand this behaviour by noting that for very small
spin-flip strength a small current can flow through the wire by
the electrons whose spins flip once. These few electrons travel
almost uncorrelated through the wire resulting in a full
Poissonian noise. For a finite $\lambda$ the value of $F$ strongly
depends on the polarization. While for $p_N=p_W=0$ the Fano factor
decreases with $\lambda$ and tends to the normal value Eq.
(\ref{ffn}) in the limit $\lambda \gg 1$, it has a nonmonotonic
variation for finite $p_N$ and $p_W$. The Fano factor at high
$\lambda$ increases above the normal value and reaches a value
which is determined by the polarization.  By approaching to the
limit of perfect DOS polarization, $p_N \rightarrow 1$, the Fano
factor tends to one. The nonmonotonic dependence on $\lambda$ is
the result of the interplay between the spin-flip process and the
DOS polarization. While the former is dominant at $\lambda \ll 1$
and increases the mean number of electrons whose spin flips once,
the later imposes a restriction on the spin-flip current at higher
(finite) $\lambda$s. Thus $F$ develops a minimum  at a $\lambda $
which depends on $p_N$. It is interesting to note that in contrast
to the mean current which can be modulated equivalently by varying
$p_N$ and $p_W$, the Fano factor $F$ is mainly affected by the
variation of $p_N$ rather than $p_W$: The Fano factor for
different $p_N$ are well separated by varying slightly with $p_W$.
Thus the shot noise measurement can be used to determine  the
intrinsic and the extrinsic contributions to the current
polarization.
\par
Let us now analyze the shot noise in the case of the P
configurations. Figs. \ref{hzfig3}a and \ref{hzfig3}b illustrate
the behavior of the shot noise in PP and PA configurations
respectively. In the PP configuration the Fano factor has a weak
variation with $\lambda$. But it varies considerably by increasing
the polarization of the conductivity. Note that $p_{\sigma}$ given
in Eq. (\ref{psigma}) is always close to $p_{W}$ except when $p_N
\rightarrow 1$ for which $p_{\sigma}$ approaches one independently
of $p_W$. The dominant dependence comes from the variation of
$p_W$ and the $p_N$-dependence is weak. In contrast to the case of
the A configuration the shot noise changes with the relative sign
of $p_N$ and $p_W$ for the P configurations. In the PA
configuration for $p_N$ approaching unity $F$ become constant
($1/2$) independent of $\lambda$.
\par
In conclusion we have presented a semiclassical Boltzmann-Langevin
theory of shot noise in a fully ferromagnetic spin-valve
consisting of a diffusive ferromagnetic wire connected by tunnel
contacts to two perfectly polarized half-metallic ferromagnetic
electrodes. We have shown that the shot noise can distinguish the
intrinsic DOS polarization contribution from the extrinsic one
induced by the scattering from the normal and the magnetic
disorders. While the shot noise for an antiparallel configuration
of the magnetization vectors in the electrodes has a weak
dependence on the extrinsic impurity polarization, it is sensitive
to the intrinsic DOS polarization. At a finite spin-flip
scattering rate the shot noise has been shown to increase
substantially above the unpolarized value and reaches the full
Poissonian value for a perfectly polarized DOS. In the parallel
configuration we have found that the shot noise is sensitive to
the relative sign of the intrinsic and the extrinsic
polarizations. Our result reveals importance of the shot noise as
a possible probe to distinguish the intrinsic and the extrinsic
anomalous Hall effects in ferromagnets.

\acknowledgments We acknowledge fruitful discussions with G. E. W.
Bauer, A. Brataas and A. Fert.

\end{document}